\documentclass[prl,showpacs,twocolumn]{revtex4}%
\usepackage{amsmath}
\usepackage{amssymb}
\usepackage{graphicx}
\usepackage{dcolumn}
\usepackage{bm}
\usepackage{amsmath}
\usepackage{graphicx}
\usepackage{amsfonts}
\usepackage{subfigure}
\usepackage{graphicx}
\usepackage{float}

\usepackage{amssymb}%

\begin{document}
\title{Entanglement of light-shift compensated atomic spin waves with telecom light}
\date{\today }
\author{Y. O. Dudin}
\author{A. G. Radnaev}
\author{R. Zhao}
\author{J. Z. Blumoff}
\author{T. A. B. Kennedy}
\author{A. Kuzmich}
\address{School of Physics, Georgia Institute of Technology, Atlanta,
  Georgia 30332-0430}

\pacs{42.50.Dv,03.65.Ud,03.67.Mn}

\begin{abstract}
  Entanglement of a 795 nm light polarization qubit and an atomic Rb spin wave qubit for a storage time of 0.1 s is observed by measuring the violation of Bell's inequality ($S=2.65 \pm 0.12$). Long qubit storage times are achieved by pinning the spin wave in a 1064 nm wavelength optical lattice, with a magic-valued magnetic field superposed to eliminate lattice-induced dephasing. Four-wave mixing in a cold Rb gas is employed to perform light qubit conversion between near infra red (795 nm) and telecom (1367 nm) wavelengths, and after propagation in a telecom fiber, to invert the conversion process. Observed Bell inequality violation ($S=2.66 \pm 0.09$), at 10 ms storage, confirms preservation of memory/light entanglement through the two stages of light qubit frequency conversion.
\end{abstract}

\maketitle

Future quantum information processing systems will rely on the ability to
generate, distribute and control elementary entanglement processes across continental distances. Besides offering fundamentally more secure ways to communicate, quantum networks may provide the structure for distributed quantum computation. Large-scale quantum networks necessarily require mitigation of exponential photon transmission losses, by using compatible quantum memory elements and so-called quantum repeater protocols \cite{briegel,duan}.  Compatibility involves storing and retrieving quantum information and transmitting the latter optically, in the case of fiber-based networks, in the telecom wavelength range where absorption is minimized \cite{chaneliere}. Unfortunately, typical atomic ground-state electronic transitions suitable for quantum information applications lie outside the telecom window \cite{matsukevich,matsukevich1,matsukevich2,zhao,chen,chen1,rosenfeld,simon,blinov,togan}. The entanglement distribution rate of a network also depends critically on the quantum memory lifetime of the storage elements; memory lifetimes of a second or longer may be desirable \cite{briegel,duan,collins,jiang,sangouard}. The combined attributes of telecom wavelength light and a long-lived quantum memory are therefore essential for fiber-based quantum networks \cite{chaneliere}.

Previously, the entanglement of near-infra red (NIR) light at 795 nm with an atomic spin wave for a 3.3 ms storage period was reported. The memory time was limited by inhomogeneous light shifts in the optical lattice used to eliminate motional dephasing on a sub-ms time-scale \cite{dudin}. Two orthogonal components associated with the
$m=\pm1\leftrightarrow m^{\prime} =\mp1$ coherences of a single mode spin-wave were used to
encode the long-lived atomic qubit. Observations of quantum correlations of a long-lived ($\sim 0.1$ s lifetime) memory with a NIR field, and with a telecom field for 11 ms storage time, were very recently reported \cite{radnaev}.  In this work ac Stark decoherence was removed by a new two-photon laser compensation technique.

\begin{figure*}[tbh]
\includegraphics[scale=0.95]{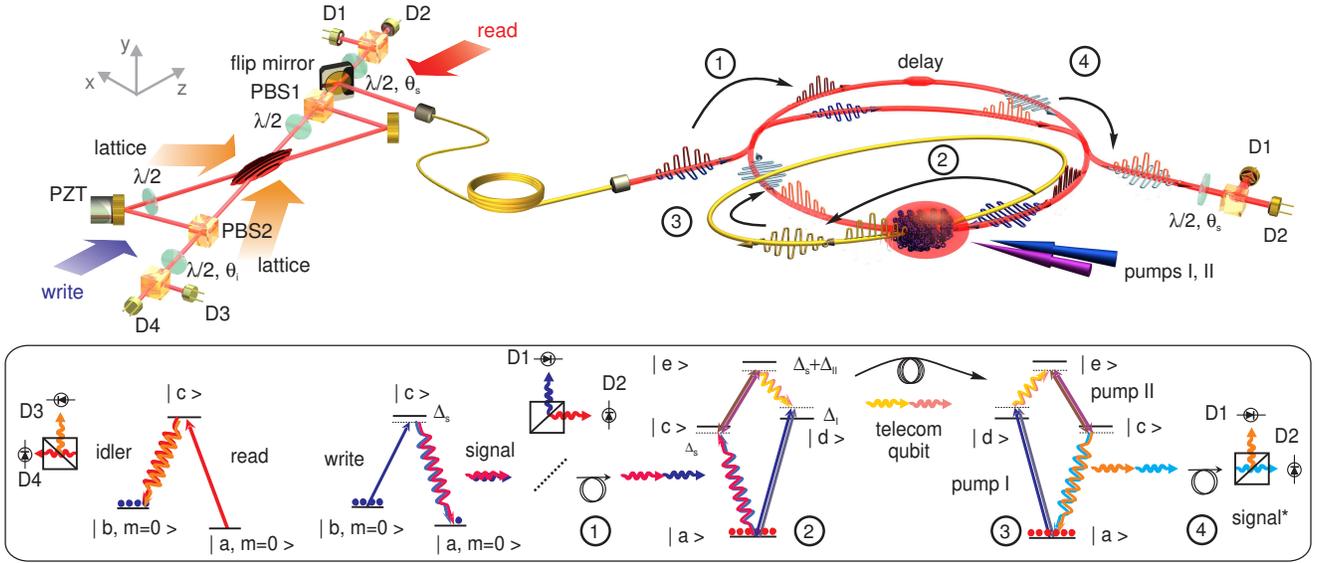}
\caption{Left side: A write laser (270 $\mu$m mode waist) generates spin-waves in atomic $^{87}$Rb,
confined in a 1-D lattice with magnetically-compensated clock transition light shifts. Trap depth $U_0 = 56$ $\mu$K, trap frequencies ($\omega_x/(2\pi), \omega_y/(2\pi), \omega_z/(2\pi)$) = (8100,116,10) Hz.
The experimental protocol is based on a sequence of write/clean pulses, terminated by photodetection of the signal field at D1 or D2 \cite{duan,matsukevich2}. After a storage period, the stored spin-wave qubit is converted by the read laser to idler field qubit and polarization measurement of the latter is performed.
Signal and idler fields intersect at the center of the trap at an angle
of $\pm$ 0.9$^{\circ}$ with respect to the z-axis, and have waist size of 120 $\mu$m. Two signal (idler) paths are overlapped on PBS1 (PBS2). The interferometric path length difference is stabilized so that the signal-idler polarization state at the output of PBS1 and PBS2 has the form $\propto (|H\rangle_s|H\rangle_i + |V\rangle_s|V\rangle_i)$. An auxiliary laser at 766 nm (not shown), intensity-stabilized and frequency-locked to the potassium $D_2$ line, is used for that purpose.
Right side: Successive frequency down- and up- conversion of the signal field qubit is realized by
four-wave mixing in cold $^{87}$Rb. A polarizing beamsplitter separates the H- and V- components of the signal field and the latter is delayed by an optical fiber (step 1).
In step 2, the write signal and pump fields generate the telecom signal ($e\rightarrow d$ transition), which is directed through a 100 m standard
telecommunication fiber back to the atomic sample. In step 3 the telecom signal is upconverted to
NIR signal ($a\leftrightarrow c$ transition). After its two polarization components are temporally overlapped (step 4) using the same interferometric arrangement used to separate the incoming NIR signal, a polarization measurement is performed. High-efficiency detection is achieved by the Si single photon detectors, D1$^{\prime}$ and D2$^{\prime}$. The inset shows the $\Lambda$-type atomic levels used for the DLCZ scheme (left) and the cascade configurations used for wavelength conversion (right): $|a\rangle=|5S_{1/2}\ F=1\rangle$, $|b\rangle=|5S_{1/2}\ F=2\rangle$, $|c\rangle=|5P_{1/2}\ F=2\rangle$,
$|d\rangle=|5P_{3/2}\ F=2\rangle$, $|e\rangle=|6S_{1/2}\ F=1\rangle$, $\mathrm{\Delta_s=-2\pi \times 17}$ MHz, $\mathrm{\Delta_{I}=2\pi \times 41}$ MHz,
$\mathrm{\Delta_{II}=2\pi \times 6}$ MHz.}
%\label{fig:setup}
\end{figure*}

Here we report measurement of entanglement between an atomic spin-wave memory qubit and a telecom field qubit, at a storage time of 10 ms. As in Ref. \cite{radnaev}, high-efficiency, low-noise wavelength conversion between NIR and telecom fields in an optically-thick, cold Rb gas is at the core of our experimental protocol. The memory qubit employs two spatially distinct spin-waves, both based on the $m=0 \leftrightarrow 0$ clock transition between hyperfine ground levels. The qubit spin-wave, pinned by an optical lattice to reduce motional dephasing, is lifetime-enhanced by mixing the clock states using a bias magnetic field to offset inhomogeneous, differential ac-Stark shifts \cite{lundblad,dudin1}. Our Bell inequality violation measurements indicate the memory qubit is entangled with the NIR qubit for as long as 0.1 s, a 30-fold improvement on the results of Ref. \cite{dudin}.

We now describe light-shift-compensated memory qubit, shown in the left panel of Fig. 1. A magneto-optical trap (MOT) captures $^{87}$Rb atoms from a background vapor. After compression and cooling the atoms are transferred to a one dimensional
optical lattice formed by interfering two
circularly polarized 1064 nm beams intersecting at an angle of 9.6$^\circ$ in the horizontal plane (a 6.3 $\mu$m lattice period). The waists of the lattice beams are $\approx 0.2$ mm and their total power is typically
12 W. The cigar shaped atomic cloud has $1/e^2$ waists of 0.15 mm and 1 mm, respectively, and contains about $10^7$ atoms. After loading, the atoms are prepared in the $|+\rangle \equiv |5^2S_{1/2}, F = 2, m = 0 \rangle$ state
by means of optical pumping.
The magnetic field is set to the ``magic" value (4.2 G for our 1064 nm wavelength lattice \cite{dudin1}) to equalize the ac-Stark shifts of the clock states.

Raman scattering of a weak linearly
polarized {\it write} laser field transfers a fraction of atomic population into the $|-\rangle \equiv |5^2S_{1/2}, F = 1,m = 0\rangle$ clock state generating a
signal field offset in frequency from the \textit{write} field by $ 6.8$ GHz, the $^{87}$Rb ground-level hyperfine splitting. The memory qubit is based on two spatially distinct spin-waves \cite{matsukevich,chen1}. The interferometric arrangement for the associated signal and idler field modes is shown in Fig. 1. Two Gaussian modes of the signal field centered at angles $\pm 0.9^{\circ}$  to the \textit{write} field, $\bf{k_w}$, direction are combined on a polarizing beamsplitter (PBS1), and directed to either the polarization measurement setup involving detectors D1 and D2 for spin-wave memory/NIR field entanglement analysis, or, via an optical fiber, to the wavelength conversion setup to generate memory/telecom field entanglement. The signal field photoelectric detection probability per experimental trial is in the range $10^{-4}-10^{-3}$ and can be varied by adjusting the {\it write} field power/detuning.

The detection of the signal photon implies a momentum change
$\hbar(\bf{k_w-k_s})$ of the atoms,
where $\bf{k_s}$ is the signal field
wavevector. An atom at
position $\bf{r_{\mu}}$ contributes to the spin-wave excitation with a phase factor $e^{-i\bf{(k_w-k_s)}\cdot
\bf{r_{\mu}}}$; the atom experiences additional phase modulation due to the local value of the differential ac-Stark shift. The collective atomic excitation, imprinted with
this phase grating, is the {\it write} spin wave. The spin wave
coherence is essential for efficient coupling to a single
spatial electromagnetic field mode in the memory retrieval stage.

The retrieval, or {\it
read} process, is performed after a defined storage period. A \textit{read} field pulse converts
the stored spin-wave excitations into an idler field by Raman scattering, Fig. 1. The idler emission is collected in the two Gaussian spatial modes of the detected signal field, but with opposite propagation directions ($\bf{k_i} \approx -\bf{k_s}$). The two idler modes are combined on a polarization beamsplitter (PBS2) and directed to detectors D3 and D4 for polarization analysis.

We have previously observed quantum memory lifetime (1/e) of 0.3 s for a single spin-wave aligned such that the vector $\bf{k_w-k_s}$ is directed along the axis of our 1-D lattice \cite{dudin1}.
 In the arrangement shown in Fig. 1, the vectors $\bf{k_w-k_{s_j}}$  for each of the two qubit spin-waves, $j=1,2$, are at angles $\phi \approx \pm 0.5^{\circ}$ to the lattice axis (x-axis). We have measured the efficiency of light storage and retrieval as a function of storage time to characterize the coherence properties of the two spin waves. Storage of a classical field is achieved by counterpropagating the light in the signal field fiber, while the \textit{read} light is used in the role of control field. In this geometry the stored spin waves are of the same 50 $\mu$m period and have a similar spatial envelope as the single excitations created via the DLCZ protocol. The spin-wave period is much longer than that of the lattice, resulting is strong suppression of motional dephasing. We observe that the storage lifetime is rather sensitive to the alignment of the control and signal fields relative to the optical lattice and to the size of the lattice beams. Even when the interferometer is optimally aligned, the observed storage lifetime of $\approx 0.2$ s is shorter than the $\approx 0.3$ s reported in Ref. \cite{dudin1}. The time-dependence of the retrieval efficiency exhibits deviations from the near exponential decay observed in Ref. \cite{dudin1}. We attribute these observations to atomic motion along the z-axis, which, due to the non-zero value of $\phi$, partially dephases the spin-wave coherence.

The polarization states of both the signal and idler fields are
measured using a polarizing beam splitter and two single
photon detectors, D1, D2 for the signal and D3, D4 for the idler
(additional technical details are given in Ref. \cite{dudin}).
We denote the joint signal-idler detection rates between detector $Dn $, $n =1,2$ and detector $Dm$, $m =3,4$ by  $ C_{nm}\left(\theta _s,\theta _i\right)$.
Here $\theta_s$ and $\theta_i$ are the angles through which the polarization is rotated by the respective half-waveplates.
The rates
$C_{ij}\left({\theta_s},\theta _i\right)$, $i=1,2,j=3,4$ exhibit
sinusoidal variations as a function of the waveplates' orientations. In order to account for unequal efficiencies of the
detectors (D1, D2) and (D3,D4), each correlation measurement
includes flipping the polarization of both signal and idler fields by $\pi/2$
within the data set. We check for Bell's inequality
violation $|S|\leq 2$  by measuring polarization correlations
between signal and idler fields at certain canonical angles, where
$S = E(\pi/4, \pi/8) + E(0, \pi/8) + E(0, -\pi/8) -
E(\pi/4, -\pi/8)$; here the correlation function $E(\theta_s,
\theta_i)$ is given by \cite{bell,clauser}
\begin{equation*}
  \frac{C_{13}(\theta_s,\theta_i) + C_{24}(\theta_s,\theta_i) -
    C_{14}(\theta_s,\theta_i) - C_{23}(\theta_s,\theta_i)}
  {C_{13}(\theta_s,\theta_i) + C_{24}(\theta_s,\theta_i) +
    C_{14}(\theta_s,\theta_i) + C_{23}(\theta_s,\theta_i)}.
    \label{corr_func}
\end{equation*}

\begin{figure}[tbh]
  \centering
  \includegraphics[width=3.1in]{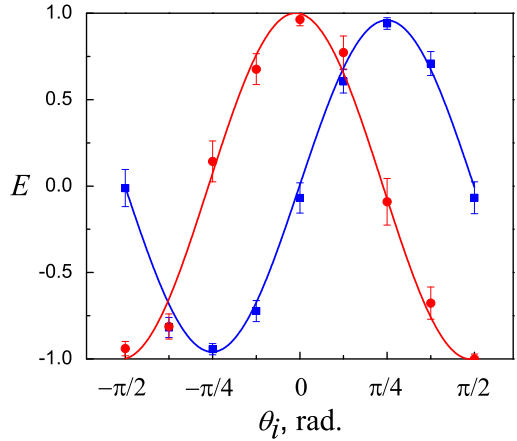}
  \vspace{-0.2cm}
  \caption{Measured values of the correlation function $E(\theta _s, \theta _i)$ as a function of $\theta_i$ for 1 ms storage. Circles are for $\theta_s=0$, squares are for $\theta_s=\pi/4$. The curves are sinusoidal fits to the data.}
  \label{fig.bell_data}
\end{figure}

The measured sinusoidal variation of $E(\theta_s, \theta_i)$ as a
function of $\theta_i$ for fixed $\theta_s$ is shown in Fig. 3. In Table 1
we list the values of E and the Bell parameter S for two values of storage time: 1 ms and 0.1 s. The measured idler photodetection probabilities for the data in Table 1 are $1.7\%$ at 0.1 s and $4\%$ at 1 ms. Normalized by the passive transmission and detection losses (0.25), the intrinsic atomic excitation retrieval efficiency after 0.1 s storage is $7\%$ ($16\%$ at 1 ms). We observe different behavior of the retrieval efficiencies as a function of storage time for coherent light and the DLCZ protocol. This could be caused by differences in the spatial mode structure of the retrieved excitations, as we have observed previously \cite{radnaev}.

To convert the signal photons produced by the \textit{write} process into light of telecom wavelength we employ the diamond configuration of atomic levels shown in the inset to Fig. 1 \cite{chaneliere,willis,jen}. A cigar-shaped ($\varnothing\sim$2 mm, $\mathrm{L\sim}$6 mm) sample of $^{87}$Rb gas is prepared, in level $|a\rangle$ with optical depth $\sim 150$, in an extended dark magneto-optical trap (EDMOT). Additional details are given in Ref. \cite{radnaev}.

\begin{table}[bht]
\caption{\label{tab:table1} Measured correlation function $E(\theta_s,
  \theta _i)$ and $S$ for 1 ms and 100 ms storage time. These are based on 582 events and 1001 events, respectively.}
\begin{ruledtabular}
\begin{tabular}{ccccc}
  $$ & $$& $~~~~~~~~~~~~~~~~~~~~~~~~~~~~~~~~~E(\theta_s, \theta _i)$  \\
  $\theta_s$ & $\theta _i $     &  1 ms      &              &0.1 s  \\
  \hline
  $\pi/4$  & $-\pi/8$           &  $-0.78 \pm 0.05$      &                 &   $-0.65  \pm 0.05$   \\
  $\pi/4$  & $\pi/8$            &  $0.71 \pm 0.07$       &                 &   $0.67  \pm 0.05$   \\
  $0$  & $-\pi/8$               &  $0.75 \pm 0.05$        &                 &    $0.66  \pm 0.05$   \\
  $0$  & $\pi/8$                &  $0.66 \pm 0.06$         &                 &     $0.68 \pm 0.05$   \\
           &                    &  $S=2.90 \pm 0.12$         &                   & $S=2.66 \pm 0.09$          \\
         \end{tabular}
\end{ruledtabular}
\end{table}

We have previously demonstrated high-
efficiency, low-noise telecom wavelength conversion for single-photon light fields \cite{radnaev}. However, the high (up to $65\%$) efficiency of frequency conversion is only observed for light that is copolarized with the telecom pump. The orthogonal polarization component of the signal field is largely absorbed in the atomic gas. Here we achieve polarization-stable wavelength conversion by splitting the two polarization components of the optical field and delaying the (vertically polarized) V-component by 235 ns (Fig. 1). By switching the polarization of both pump fields to orthogonal settings while the V-component interacts with the atoms we achieve high transparency and high ($54\%$) conversion efficiency for both (horizontally polarized) H- and V- components. The conversion efficiency is lower than the $65\%$ observed for a single polarization component, as a result of the limited power available for pump II and the denser atomic sample used to compensate for the undesirable effects of hyperfine optical pumping.

\begin{figure}[thb]
  \centering
  \includegraphics[width=3.1in]{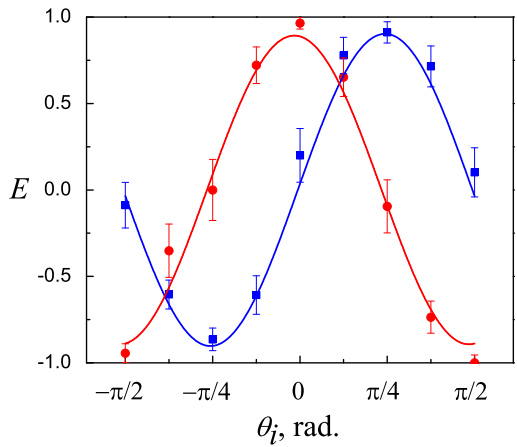}
  \vspace{-0.2cm}
  \caption{Measured correlation function $E(\theta _s, \theta _i)$ when 795 nm signal field is first converted to telecom wavelength, passed through 100 m telecom fiber, and converted back to 795 nm field, for 1 $\mu$s storage. Circles are for $\theta_s=0$, squares are for $\theta_s=\pi/4$.The curves are sinusoidal fits to the
data.}
  \label{fig.bell_data}
\end{figure}

\begin{table}[htb]
\caption{\label{tab:table1} Measured correlation function $E(\theta_s,
  \theta _i)$ and $S$ when 795 nm signal field is first converted to telecom wavelength, passed through 100 m telecom fiber, and converted back to 795 nm field. Storage times are 1 $\mu$s and 10 ms, based on 986 and 667 events, respectively.}
\begin{ruledtabular}
\begin{tabular}{ccccc}
   $$ & $$& $~~~~~~~~~~~~~~~~~~~~~~~~~~~~~~~~~E(\theta_s, \theta _i)$  \\
  $\theta_s$ & $\theta _i $     &  1 $\mu$s&              &10 ms  \\
  \hline
    $\pi/4$  & $-\pi/8$           &  $-0.54  \pm 0.05$ &             &  $-0.61  \pm 0.06$   \\
  $\pi/4$  & $\pi/8$                &  $0.65  \pm 0.05$ &                    &  $0.72  \pm 0.06$   \\
  $0$  & $-\pi/8$                 &  $0.78  \pm 0.04$    &                   &  $0.75  \pm 0.05$   \\
  $0$  & $\pi/8$                    &  $0.58 \pm 0.05$    &                &  $0.56 \pm 0.07$   \\
 &                  & $S=2.55 \pm 0.10$ &                 & $S=2.64 \pm 0.12$          \\
         \end{tabular}
\end{ruledtabular}
\end{table}

The level of dark counts of commercially available
InGaAs/InP single photon detectors for telecom wave-
lengths is too high for our purposes. Instead, we employ
a reverse wavelength conversion sequence, from 1367 nm
to 795 nm, Fig. 1. The H- and V- components of the regenerated NIR signal field are subsequently recombined. A half-wave plate, a
polarizing beam splitter and two Si single photon detectors D$1^{\prime}$ and D$2^{\prime}$ complete the measurement of the
polarization state of the telecom signal field qubit. The combined transmission measured for the NIR signal qubit from
the input of the signal field fiber to detectors D$1^{\prime}$ and D$2^{\prime}$ is $7.5(5)\%$, with a factor 0.25 contribution from passive optical elements and fiber coupling losses (0.8 for both telecom and NIR fields).
The interferometric configuration for the signal field sketched
in the right panel of Fig. 1 is designed to cancel out phase
fluctuations occurring on a time scale slower than  1 $\mu$s, so that active stabilization of the interferometer is not required.
The ellipticity of the signal field, acquired due to birefringence of the optical fibers, is removed by adjusting the phase difference between the H- and V- components of the pump II pulses employed for the upconversion.
With coherent light input the polarization state is preserved after transmission and conversion with a power contrast of greater than 100 to 1.
We have not observed a noticeable drift of the polarization fringes on a timescale of two days, for either coherent
laser or signal field inputs to the wavelength
conversion set-up.

The measured sinusoidal variation of $E(\theta_s, \theta_i)$ as a
function of $\theta_i$ for fixed $\theta_s$ is shown in Fig. 4, for a short period of 1 $\mu$s. In Table 2
we give the measured values of $E$ and the Bell parameter S for 1 $\mu$s and for 10 ms storage. The measured idler photodetection probabilities for the data in Table 2 are $5\%$ at 1 $\mu$s and $2.8\%$ at 10 ms. The data was taken with larger lattice beams (0.26 mm waist size instead of 0.2 mm for the data without wavelength conversion) resulting in a revival behavior of the retrieval efficiency in the 30-100 ms region; we attribute this to atomic motion along the long axis of the trap.

In summary, we have confirmed preservation of memory/light entanglement through two stages of telecom wavelength conversion, for 10 ms storage. We have also observed entanglement of a spin-wave produced in a 1-D optical lattice and a NIR light field for a 0.1 s storage time. The storage time is limited by atom motion along the principal trap axis, and could be further increased by confining atoms in 2-D or 3-D lattice geometries.

We thank S. D. Jenkins and D. N. Matsukevich for discussions. This work was supported by the Air Force Office of Scientific Research, the Office of Naval Research, and the National Science Foundation.


\begin{thebibliography}{99}
\bibitem{briegel} H.-J. Briegel {\it et al.}, {Phys. Rev. Lett.} \textbf{81}, 5932 (1998).
\bibitem{duan} L.-M. Duan, M. D. Lukin, J. I. Cirac, and P. Zoller, {Nature} \textbf{414}, 413 (2001).
\bibitem{chaneliere} T. Chaneli\`{e}re {\it et al.}, {Phys. Rev. Lett.} \textbf{96}, 093604 (2006).
\bibitem{matsukevich} D. N. Matsukevich and A. Kuzmich, {Science} {\bf 306}, 663 (2004).
\bibitem{matsukevich1} D. N. Matsukevich {\it et al.} {Phys. Rev. Lett.}
\textbf{95}, 040405, (2005).
\bibitem{matsukevich2} D. N. Matsukevich {\it et al.}, {Phys. Rev. Lett.} \textbf{97}, 013601 (2006).
\bibitem{rosenfeld}  W. Rosenfeld {\it et al.}, {Phys. Rev. Lett.} \textbf{101}, 260403 (2008).
\bibitem{simon} J. Simon, H. Tanji, S. Ghosh, and V. Vuletic, {Nature Physics} \textbf{3}, 765 (2007).
\bibitem{chen1} S. Chen et al, Phys. Rev. Lett. {\bf 99}, 180505 (2007).
\bibitem{chen} Y.A. Chen {\it et al.}, {Nature Physics} \textbf{4}, 103 (2007).
\bibitem{zhao} R. Zhao {\it et al.}, {Nature Physics} \textbf{5}, 100 (2009).
\bibitem{blinov} D. N. Matsukevich, P. Maunz, D. L. Moehring, S. Olmschenk, and C. Monroe, {Phys. Rev. Lett.} {\bf 100}, 150404 (2008).
    \bibitem{togan} E. Togan {\it et al.}, Nature {\bf 466}, 730 (2010).
\bibitem{collins} O. A. Collins, S. D. Jenkins, A. Kuzmich, and T. A. B. Kennedy, {Phys. Rev. Lett.}
{\bf 98}, 060502  (2007).
\bibitem{jiang} L. Jiang, J. M. Taylor, and M. D. Lukin, {Phys. Rev. A} \textbf{76}, 012301 (2007).
\bibitem{sangouard}  N. Sangouard {\it et al.}, {Phys. Rev. A} \textbf{77}, 062301 (2008).
\bibitem{dudin} Y. O. Dudin {\it et al.}, {Phys. Rev. Lett.} {\bf 103}, 020505 (2009).
\bibitem{radnaev} A. G. Radnaev {\it et al.}, {Nature Physics}, in press.
\bibitem{lundblad} N. Lundblad, M. Schlosser, and J. V. Porto, {Phys. Rev. A} \textbf{81}, 049904 (2010).
\bibitem{dudin1} Y. O. Dudin, R. Zhao, T. A. B. Kennedy, and A. Kuzmich, {Phys. Rev. A} \textbf{81}, 041805 (2010).
\bibitem{bell} J. S. Bell, {Physics} \textbf{1}, 105, (1964); J. S. Bell, {Rev. Mod.
Phys.} \textbf{38}, 447 (1966).
\bibitem{clauser} J. F. Clauser, M. A. Horne, A. Shimony,
and R. A. Holt, Phys. Rev. Lett. \textbf{23}, 880 (1969).
\bibitem{willis} R. T. Willis {\it et al.}, {Phys. Rev. A} {\bf 79}, 033814 (2009).
\bibitem{jen} H. H. Jen and T. A. B. Kennedy, {Phys. Rev. A} {\bf 82}, 023815 (2010).
\end{thebibliography}
\end{document}